\documentclass[preprint]{revtex4}

\usepackage[english]{babel}
\usepackage[dvips]{color}
\usepackage{graphicx}
\usepackage{epsfig}
\usepackage{float}

\usepackage{amsfonts,amsmath}

\usepackage{hyperref}

\newcommand{\dd}{{\rm d}}
\newcommand{\E}{{\cal E}}
\newcommand{\U}{{\cal U}}
\newcommand{\N}{{\cal N}}

\newcommand{\C}{{C}}

\begin{document}

\title{\bf{Dynamics of a 1-D model for the emergence of the
plasma edge shear flow layer with momentum conserving Reynolds
stress}}

\author{I. Calvo\vskip0.3cm}
\affiliation{Laboratorio Nacional de Fusi\'on\\Asociaci\'on
EURATOM-CIEMAT\\28040 Madrid, Spain}
\author{B. A. Carreras\vskip0.3cm}
\affiliation{BACV Solutions Inc.\\Oak Ridge, TN 37830, U.S.A.\vskip2cm}

\begin{abstract}
A one-dimensional version of the second-order transition model based
on the sheared flow amplification by Reynolds stress and turbulence
supression by shearing is presented. The model discussed in this paper
includes a form of the Reynolds stress which explicitly conserves
momentum. A linear stability analysis of the critical point is
performed. Then, it is shown that the dynamics of weakly unstable
states is determined by a reduced equation for the shear flow. In the
case in which the flow damping term is diffusive, the stationary
solutions are those of the real Ginzburg-Landau equation.
\end{abstract}

\maketitle

\section{Introduction}

The existence of a shear flow layer at the tokamak edge and in ohmic
discharges has been known for a long time \cite{Ritzetal,ZweGou}. It
was later found that similar edge shear flow layers existed in other
confinement devices. This seems to be a generic feature of confined
plasmas. In the last years, a great deal of attention has been
directed to the formation of shear flow layers and the corresponding
region of radial electric field gradient to understand the improved
confinement regimes. Many of the recent theoretical developments in
this direction have been focused on barrier formation \cite{Ter00} or
zonal flows \cite{DiaItoItoHahm1}. On the experimental side, much
progress has been done in the visualization of edge turbulence and
flows (\cite{Zwe04}, \cite{Zwe06}), including specific applications to
the emergence of the shear flow layer in the TJ-II stellarator
(\cite{Alo06}), in which we are especially interested.

Here, we want to turn back to the basic plasma edge shear flow
layer. In stellarators, unlike in tokamaks, one can operate at
densities for which no shear layer is present in the plasma edge,
being thus possible to study its formation.

The emergence of the plasma edge shear flow layer as the density
increases in the TJ-II stellarator \cite{AleAloBot} is shown
\cite{HidPedGarWar,PedHidCal05} to have the characteristic properties
of a second order phase transition. It is consistent
\cite{CarGarPedHid06} with a simple transition model that couples
shear flow amplification by turbulence \cite{DiaKim,CarLynGar91} with
turbulence suppression by sheared flows \cite{BigDiaTer}. The model
used in interpreting the TJ-II results is based on a transition model
\cite{DiaLiaCarTer94} initially introduced to explain the transition
from the low confinement mode (L mode) to the high confinement mode (H
mode) \cite{WagBecBehetal} in magnetically confined plasmas. This
model consists of two envelope equations for the fluctuation level and
mean poloidal flow. A later extension of the model
\cite{CarNewDiaLia94} included a third equation to account for the
pressure gradient contribution to the radial electric field. This
second model shows the existence of two critical points, the second
one causing the first order transition that has been associated with
the L to H transition. For this transition there is a hysteresis cycle
and the transition is characterized by an S-curve (see for instance
\cite{HubCarBoietal}). The first critical point leads to a second
order transition (consequently, it does not have a hysteresis cycle),
which has been identified with the emergence of the plasma edge shear
flow layer.

In this paper, we focus on this second order transition. By excluding
the diamagnetic term in the momentum balance equation, only this
transition is included in the model. This simplification is reasonable
because there is a large range of densities separating the two
critical points. For the same reason, the range of plasma parameters
considered is not yet in an L-mode regime. Therefore, we do not expect
avalanche-like transport \cite{DiaHah,NewCarDiaHah} and the transport
terms can be represented by purely diffusive terms. Concretely, we
discuss an extension to 1-D of the original model used in comparison
with the experimental data \cite{CarGarPedHid06}. In contrast with
previous 1-D extensions of the transition model
\cite{DiaLebNewCar95,delCasCar02}, here we formulate the Reynolds
stress term as a momentum conserving term. The model is defined by
three 1-D partial differential equations describing the evolution of
the turbulent fluctuation level $E$, the averaged poloidal velocity
shear $U$ and (minus) the pressure gradient $N$. It predicts a
second-order phase transition with order parameter $U$ and control
parameter $\Gamma$, the particle flux, which enters the model through
the boundary conditions. For $\Gamma$ below a critical value
$\Gamma_c$ the stable stationary solutions have $U=0$, whereas for
$\Gamma>\Gamma_c$ such solutions are unstable and undergo a transition
to states with $U\neq 0$ and reduced turbulence fluctuations.

After performing a detailed stability analysis of the model we study
two interesting special cases depending on the form of the
flow-damping term: collisional drag and collisional diffusion. For
both of them we find reduced equations describing the dynamics of
weakly unstable states. In the latter case, the reduced equation is
closely related to the Ginzburg-Landau equation for second-order phase
transitions. In particular, the stationary solutions of our equation
are exactly those of the Ginzburg-Landau one.

The paper is organized as follows:

In Section II we introduce the one-dimensional transition model with
momentum conservation. Section III is devoted to the study of the fixed
points of the model, their linear stability properties and a general
discussion of the critical conditions. In Section IV we consider the
dynamics near the critical point. Section V contains the conclusions
and an outline of future research lines.

\section{The one-dimensional transition model}

The relevant plasma edge region to which this model is applied
corresponds to $r\in[r_0,a]$, $(a-r_0)/a\approx 0.1$, where $a$ is the
minor radius of the plasma. We take the slab geometry approximation in
representing this region and use as coordinate $x:=(r-r_0)/(a-r_0)$,
so that $x\in[0,1]$.

The fields of our model will be the fluctuation level envelope
$\E:=\langle(\tilde n_k/n_0)^2\rangle^{1/2}$, the averaged poloidal
shear flow $\U:=\partial\langle V_\theta\rangle/\partial r$ and
(minus) the averaged pressure gradient $\N:=-\partial \langle
p\rangle/\partial r$, where $\langle\cdot\rangle$ denotes ensemble
average. A suitable one-dimensional generalization of the model
discussed in Ref. \cite{CarGarPedHid06} requires a form of the
Reynolds stress which conserves momentum. Using a
pressure-gradient-driven turbulence model and assuming densely packed
turbulence, a quasi-linear calculation yields the following form for
the Reynolds stress:
\begin{equation}
\langle \tilde V_x \tilde V_\theta \rangle =
\bar\alpha_3\E^2\partial_x\langle V_\theta\rangle+\bar
D_2\E^2\partial_x^3\langle V_\theta\rangle,
\end{equation}
which is very similar to the expression previously derived in
Ref. \cite{DiaRosHinMalFleSmo} in the context of zonal flow
dynamics. Notice that $\bar\alpha_3$ measures the strength of the
Reynolds stress and non-zero $\bar D_2$ is needed for the spectrum of
the instability to be bounded.

The model discussed in the present work is:
\begin{subequations}\label{eq:1dmodeloldvar}
\begin{align}
&\partial_{\bar t} \E=\gamma_0\N\E-\alpha_1\E^2-\alpha_2\U^2\E
+\partial_x\left[(\bar D_0+\bar D_1\E)\partial_x\E\right]
\label{eq:1dmodelEo}\\
&\partial_{\bar t}{\U}=-\bar\mu_1\U+\bar\mu_2\partial_x^2\U-
\bar\alpha_3\partial_x^2(\E^2\U)-
\bar D_2\partial_x^2(\E^2\partial_x^2\U)\label{eq:1dmodelSigmao}\\
&\partial_{\bar t} \N=\partial_x^2\left[(\bar D_3\E+\bar
D_4)\N\right].\label{eq:1dmodelNo}
\end{align}
\end{subequations}

Here, $\gamma_0\N$ is the linear growth rate of the characteristic
instability and $\alpha_1$ is computed from the nonlinear saturation
condition of the instability, both in the absence of sheared flow. The
$\alpha_2$ coefficient is derived from the condition of turbulence
supression by sheared flow. $\bar D_0$ and $\bar D_4$ are neoclassical
diffusivity coefficients, whereas $\bar D_1$ and $\bar D_3$ are the
coefficients of anomalous diffusivity multiplying the fluctuation
level. Finally, $\bar\mu_1$ and $\bar\mu_2$ are the coefficients of
the collisional flow-damping terms.

We can eliminate the explicit dependence on the parameters $\gamma_0$,
$\alpha_1$ and $\alpha_2$ by means of the following change of
variables:
\begin{equation}
E:=\frac{\alpha_1}{\gamma_0}\E,\quad
U:=\sqrt{\frac{\alpha_2}{\gamma_0}}\U, \quad t:=\gamma_0\bar t, \quad N:=\N,
\end{equation}
and Eqs. (\ref{eq:1dmodeloldvar}) read in terms of $E(x,t)$, $U(x,t)$
and $N(x,t)$:
\begin{subequations}\label{eq:1dmodel}
\begin{align}
&\partial_{t} E=NE-E^2-U^2E+
\partial_x\left[(D_0+D_1E)\partial_xE\right],\label{eq:1dmodelE}\\
&\partial_{t}{U}=-\mu_1U+\mu_2\partial_x^2U-
\alpha_3\partial_x^2(E^2U)-
D_2\partial_x^2(E^2\partial_x^2U),\label{eq:1dmodelSigma}\\
&\partial_{t} N=\partial_x^2\left[(D_3E+D_4)N\right],\label{eq:1dmodelN}
\end{align}
\end{subequations}
with
$D_0=\bar D_0/\gamma_0$, $D_1=\bar D_1/\alpha_1$,
$D_3=\bar D_3/\alpha_1$, $D_4=\bar D_4/\gamma_0$, $\mu_1=\bar\mu_1/\gamma_0$,
$\mu_2=\bar\mu_2/\gamma_0$,
$\alpha_3=\bar\alpha_3\gamma_0/\alpha_1^2$, $D_2=\bar
D_2\gamma_0/\alpha_1^2$. Finally, we choose the boundary conditions:
\begin{subequations}
\begin{align}
&\partial_x U(u,t)=\partial_x^3 U(u,t)=0, \ u=0,1, \ \forall t,\\
&\partial_xE(u,t)=0, \ u=0,1, \ \forall t,\\
&(D_3E+D_4)N\big|_{(0,t)}=\Gamma, \ \partial_x N\big|_{(1,t)}=0,
\end{align}
\end{subequations}
where $\Gamma$ is the particle flux and is the natural control
parameter of the model.

\section{Fixed points and linear stability analysis}

It is obvious from Eq. (\ref{eq:1dmodelSigma}) that any fixed point of
the model must satisfy $U=0$. Then, from Eq. (\ref{eq:1dmodelE}) we
find that $NE=E^2$. It only remains to use the boundary condition of
$N$ at $x=0$ and we finally have that
\begin{itemize}
\item[(i)] There always exists a fixed point
\begin{equation}
U_f=0,\
E_f(\Gamma)=N_f(\Gamma)=\frac{1}{2D_3}\left(\sqrt{D_4^2+4D_3\Gamma}-D_4\right).
\end{equation}
\item[(ii)] If $D_4\neq 0$ there exists a second fixed point (always
unstable, see below)
\begin{equation}
U'_f=0,\ E'_f=0,\ N'_f(\Gamma)=\Gamma/D_4.
\end{equation}
\end{itemize}

Let $(E_0,U_0,N_0)$ be a fixed point and linearize the equations
(\ref{eq:1dmodel}) around it:
\begin{eqnarray}\notag
&&E(x,t)=E_0+\xi_E e^{\gamma t+ikx}\ ,\cr
&&{U}(x,t)=\xi_{U} e^{\gamma t+ikx}\ ,\cr
&&N(x,t)=N_0+\xi_N e^{\gamma t+ikx}.
\end{eqnarray}
For the fixed point $(E'_f,U'_f,N'_f)$ the eigenvalue condition leads
to the dispersion relations:
\begin{eqnarray}
&&\gamma-N'_f+D_0k^2=0,\cr
&&\gamma+\mu_1+\mu_2k^2=0,\cr
&&\gamma+D_4 k^2=0.
\end{eqnarray}
It is clear that this fixed point is unstable for any $\Gamma>0$,
since the first dispersion relation gives, for $k=0$,
$\gamma=\Gamma/D_4$.

The fixed point $(E_f,U_f,N_f)$ is more interesting. In this case we
have:
\begin{eqnarray}
&&\gamma=-\mu_1-\mu_2k^2+\alpha_3E_f^2k^2-D_2E_f^2k^4\ ,\cr
&&(\gamma+E_f+(D_0+2D_1E_f)k^2)(\gamma+(D_4+D_3E_f)k^2)+D_3E_f^2k^2=0.
\end{eqnarray}
The second equation does not give solutions with $\gamma>0$. However,
the first one can yield an instability. Then, we ask under which
conditions
\begin{equation}\label{eq:gammak}
\gamma(k)=-\mu_1-\mu_2k^2+\alpha_3E_f^2k^2-D_2E_f^2k^4
\end{equation}
is positive. The neutral modes, i.e. the values of $k$ for which
$\gamma(k)=0$ are given by:
\begin{equation}
k_{\pm}^2=\frac{\alpha_3 E_f^2-\mu_2\pm\sqrt{(\alpha_3
E_f^2-\mu_2)^2-4\mu_1D_2E_f^2}}{2D_2E_f^2},
\end{equation}
which has real solutions $k_-,\ k_+$ if and only if
\begin{equation}\label{eq:necinscond}
\alpha_3E_f^2-\mu_2-2E_f\sqrt{\mu_1D_2}\geq 0.
\end{equation}

Now observe that the boundary conditions imply the quantization of
$k$. Namely,
\begin{equation}\label{eq:formofks}
k=n\pi,\ 0\leq n \in \mathbb{Z}.
\end{equation}
Therefore (\ref{eq:necinscond}) is only a necessary condition for the
existence of instabilities. In addition, there must exist some
$k=n\pi$ such that $k_- < n\pi < k_+$. The critical point is defined
by the minimum value of the flux, $\Gamma_c$, for which there exists an
unstable mode $k_c=n_c\pi$.

We define $E_c:=E_f(\Gamma_c)$. From (\ref{eq:gammak}) we find that
\begin{equation}
E_c^2=\frac{\mu_1+\mu_2k_c^2}{k_c^2(\alpha_3-D_2k_c^2)}\ .
\end{equation}
Thus, in particular, if $\alpha_3/D_2\leq\pi^2$ there are no unstable
modes, no matter how much we increase $\Gamma$. If
$\alpha_3/D_2>\pi^2$ at least $k=\pi$ can become unstable. Actually, if
$\sqrt{\alpha_3/D_2}\in(\pi,n\pi)$, there exist $n-1$ potentially
unstable modes $k=\pi,2\pi,\dots,(n-1)\pi$.

\section{Dynamics near marginal stability}

Our aim is to find approximate equations for the dynamics of
Eqs. (\ref{eq:1dmodel}) near (and above) the critical point.  To that
end we perform an expansion with parameter
\begin{equation}
\delta=\sqrt{\frac{\Gamma}{\Gamma_c}-1}
\end{equation}
for small $\delta$. Explicitly, we take:
\begin{subequations}
\begin{align}
&E=E_c+\delta^2E_2+\dots\\
&U=\delta U_1+\dots\\
&N=N_c+\delta^2N_2+\dots\\
&\Gamma=\Gamma_c(1+\delta^2)
\end{align}
\end{subequations}
(recall that $N_c=E_c$) and we perform a rescaling of the coordinates:
\begin{equation}
\bar\eta=\delta x,\quad \bar\tau=\delta^4 t.
\end{equation}

Expanding the equation of $N$ gives:
\begin{equation}\label{eq:slavingN}
N_2=E_c\left(1-\frac{D_3E_2}{D_4+D_3E_c}\right)
\end{equation}
whereas the equation for $E$ yields:
\begin{equation}
N_2-E_2-U_1^2=0
\end{equation}
and using (\ref{eq:slavingN}) we obtain for $E_2$:
\begin{equation}\label{eq:slavingE}
E_2=\frac{D_4+D_3E_c}{D_4+2D_3E_c}\left(E_c-U_1^2\right).
\end{equation}

Consequently, for weakly unstable states, the problem of studying the
dynamics of our model consists in finding an approximate equation for
the dynamics of $U$, $E$ and $N$ being determined at the end of the
day from the slaving conditions (\ref{eq:slavingE}),
(\ref{eq:slavingN}). The form of the reduced equation for $U$ is quite
different for $\mu_1\neq 0$ and $\mu_1=0$. That is why in the next
sections we study the cases $\mu_2=0$ ({\it collisional drag}) and
$\mu_1=0$ ({\it collisional diffusion}) separately. If the
collisionality at the plasma edge is high enough, the damping is
essentially diffusive and the limit $\mu_1=0$ is a good
approximation. At low collisionality the magnetic pumping dominates
and the relevant limit is $\mu_2=0$.

\subsection{Collisional drag}

\begin{figure}[H]
\begin{center}
\resizebox{5in}{!}{\includegraphics[angle=-90]{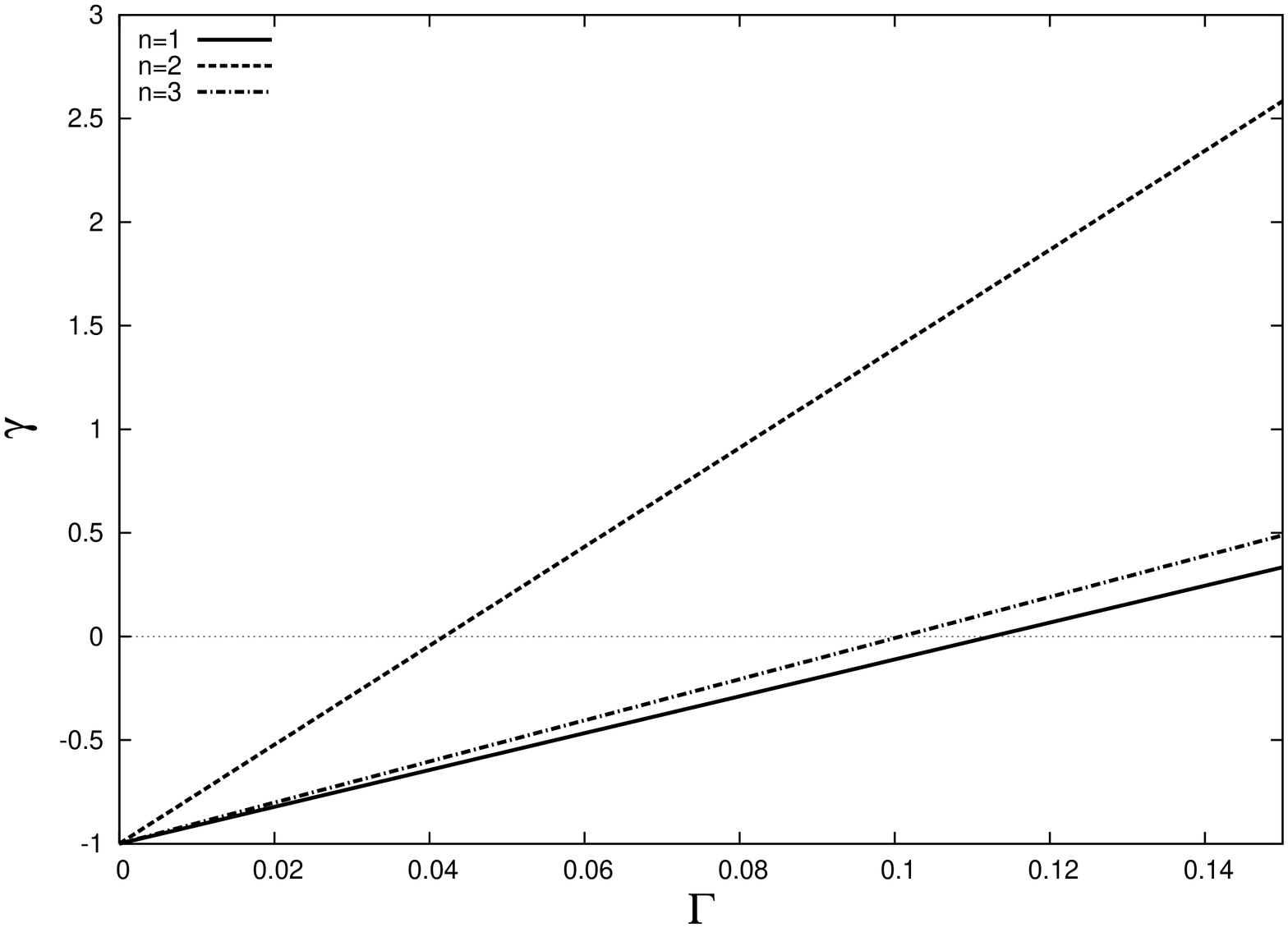}}
\end{center}
\caption{$\gamma$ as a function of $\Gamma$ near the critical
point. The values of the parameters are $\alpha_3=10^{-2}$,
$D_2=10^{-4}$, $D_3=10^{-2}$, $D_0=D_1=D_4=0$, $\mu_1=1$,
$\mu_2=0$. The critical mode is $k=2\pi$ and the critical point is
given by $\Gamma_c=0.041853$. $k=3\pi$ becomes unstable at
$\Gamma=0.1008$ and $k=\pi$ at $\Gamma=0.1124$.}
\label{fig:grate_severalns_mu2zero}
\end{figure}

\begin{figure}[H]
\begin{center}
\resizebox{5in}{!}{\includegraphics[angle=-90]{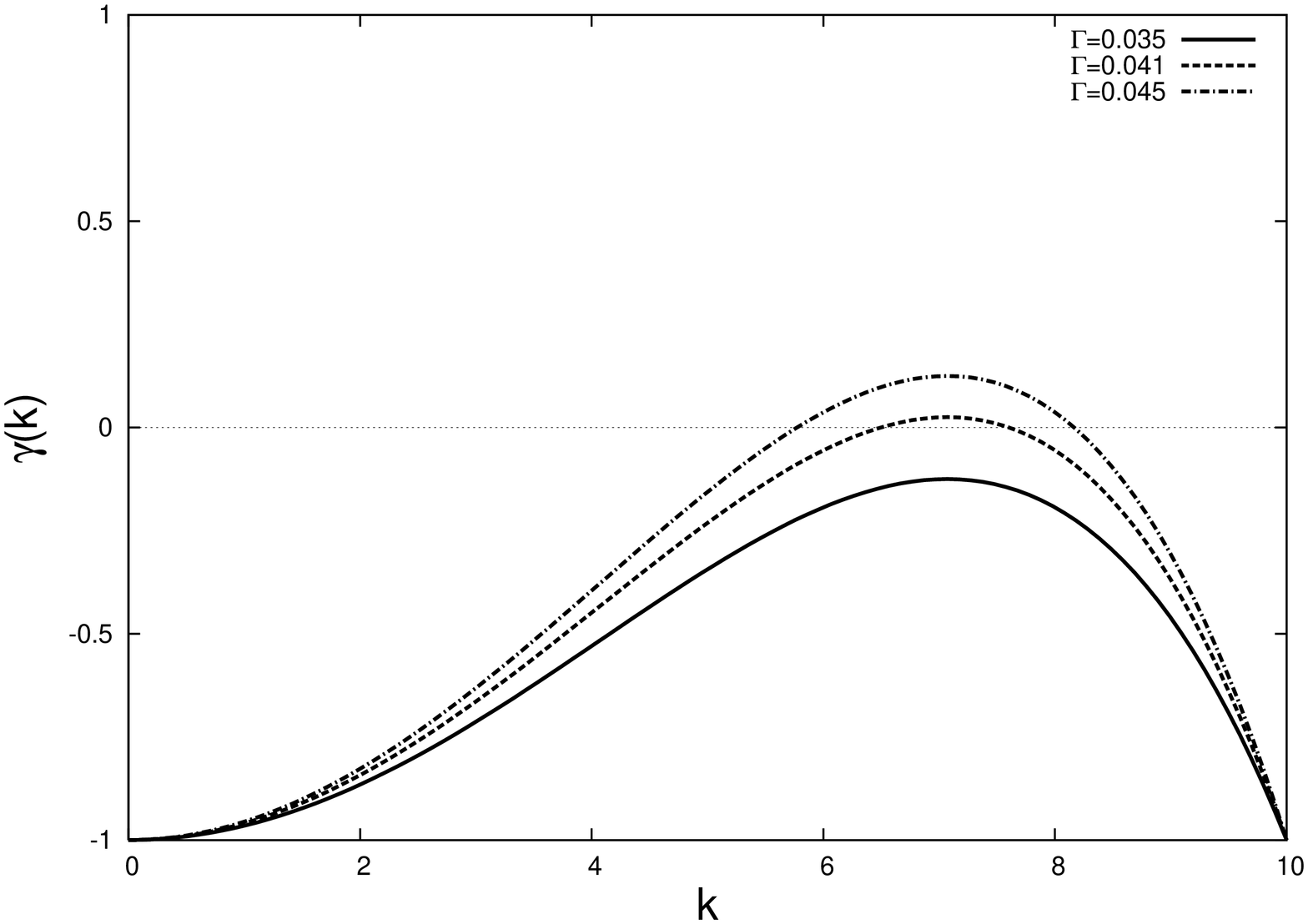}}
\end{center}
\caption{$\gamma(k)$ as a function of $k$ near the critical point. The
values of the parameters are $\alpha_3=10^{-2}$, $D_2=10^{-4}$,
$D_3=10^{-2}$, $D_0=D_1=D_4=0$, $\mu_1=1$, $\mu_2=0$. The critical
mode is $k=2\pi$ and the critical point is given by
$\Gamma_c=0.041853$.}
\label{fig:grate_mu2zero}
\end{figure}

Let us set $\mu_2=0$, $\mu_1\neq 0$. The dispersion relation reads:
\begin{equation}
\gamma(k)=-\mu_1+E_f^2k^2(\alpha_3-D_2k^2)
\end{equation}
so that
\begin{equation}
E_c:=\sqrt{\frac{\mu_1}{k_c^2(\alpha_3-D_2k_c^2)}}\ .
\end{equation}
As pointed out above, if $\alpha_3/D_2<\pi^2$ there are no unstable
modes. If $\pi^2<\alpha_3/D_2<4\pi^2$ only $k=\pi$ can be unstable and
is, of course, the critical mode. If $\alpha_3/D_2>4\pi^2$ there are
at least two modes which may become unstable and which of them is the
critical mode depends on the quotient $\alpha_3/D_2$. Hence, in
general, $k=\pi$ is not the most unstable mode (see
Figs. \ref{fig:grate_severalns_mu2zero} and \ref{fig:grate_mu2zero}).

In order to find a reduced equation for the weakly non-linear dynamics
of $U$ we expand (\ref{eq:1dmodelSigma}) keeping terms up to order
$\delta^7$:
\begin{eqnarray}\label{eq:redeqUcollision}
\delta^5\partial_{\bar\tau} U_1&=&-\mu_1\delta U_1-
\delta^3\alpha_3E_f(\Gamma)^2\partial_{\bar\eta}^2U_1+
2E_c\alpha_3\frac{D_4+D_3E_c}{D_4+2D_3E_c}\delta^5\partial_{\bar\eta}^2U_1^3
-\delta^5D_2E_f(\Gamma)^2\partial_{\bar\eta}^4U_1\cr
&+&
2D_2E_c\frac{D_4+D_3E_c}{D_4+2D_3E_c}\delta^7\partial_{\bar\eta}^2\left(U_1^2\partial_{\bar\eta}^2U_1\right),
\end{eqnarray}
where we have made use of (\ref{eq:slavingE}). Finding a reduced
equation without an explicit dependence in $\delta$ seems difficult in
this case. However, we can still obtain a useful reduced equation by
simply taking (\ref{eq:redeqUcollision}) and going back to the
original variable $U$ and coordinates $x, t$:
\begin{equation}\label{eq:redeqcollisional}
\partial_t U=-\mu_1U- E_f(\Gamma)^2\left(\alpha_3\partial_x^2U
-D_2\partial_x^4U\right)
+2E_c\frac{D_4+D_3E_c}{D_4+2D_3E_c}\left[\alpha_3\partial_x^2U^3+
  D_2\partial_x^2\left(U^2\partial_x^2U\right)\right].
\end{equation}

\begin{figure}[H]
\begin{center}
\resizebox{5in}{!}{\includegraphics[angle=-90]{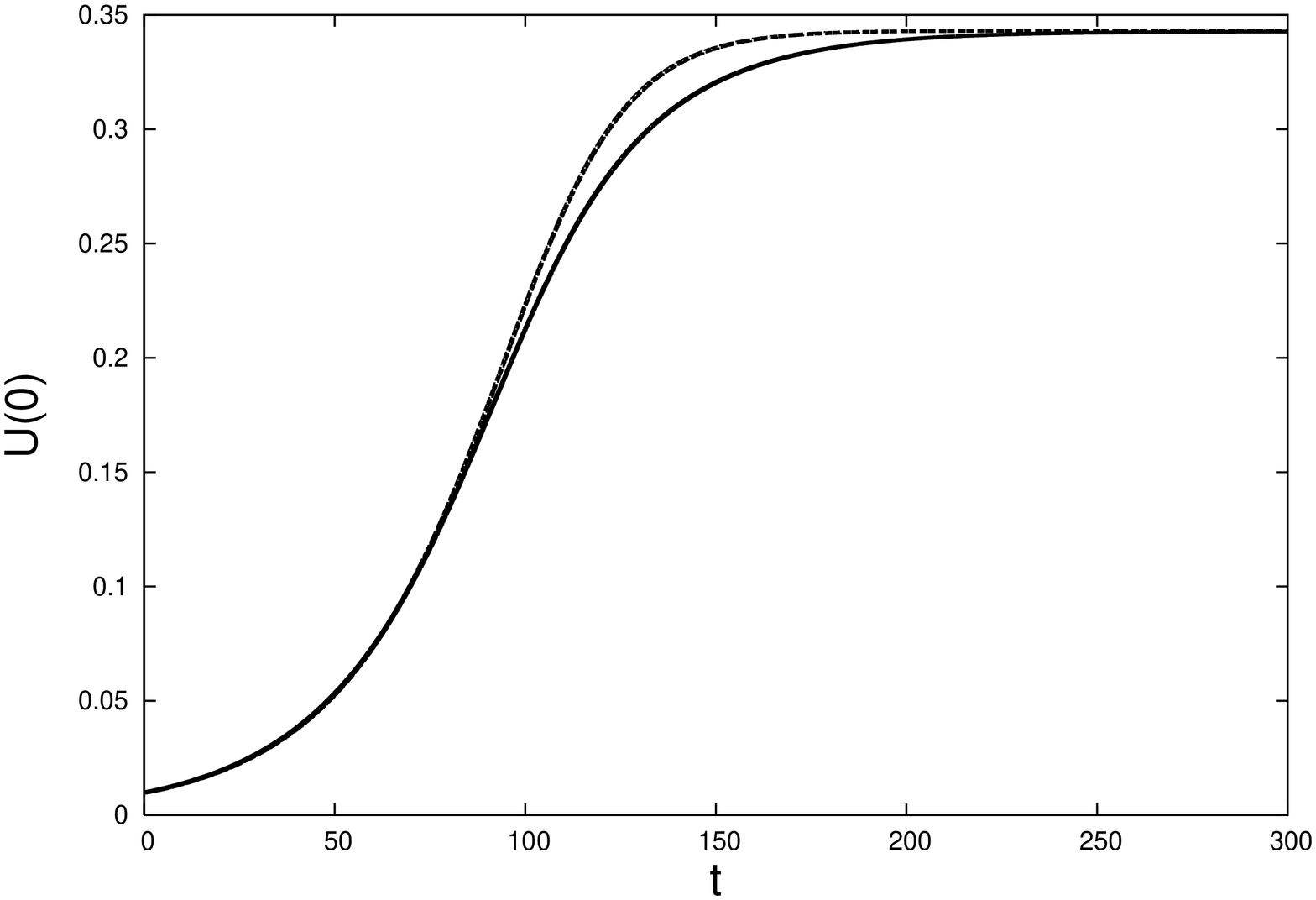}}
\end{center}
\caption{Time-evolution of $U(0)$ computed from the integration of the
original model (solid) and from the reduced equation
(\ref{eq:redeqcollisional}) (dashed). The values of the parameters are
$\alpha_3=10^{-2}$, $D_2=10^{-4}$, $D_3=10^{-2}$, $D_0=D_1=D_4=0$,
$\mu_1=1$, $\mu_2=0$, $\Gamma=0.045$. The critical point is
$\Gamma_c=0.041853$.}
\label{fig:evolutionU0collision}
\end{figure}

\begin{figure}[H]
\begin{center}
\resizebox{2.5in}{!}{\includegraphics[angle=-90]{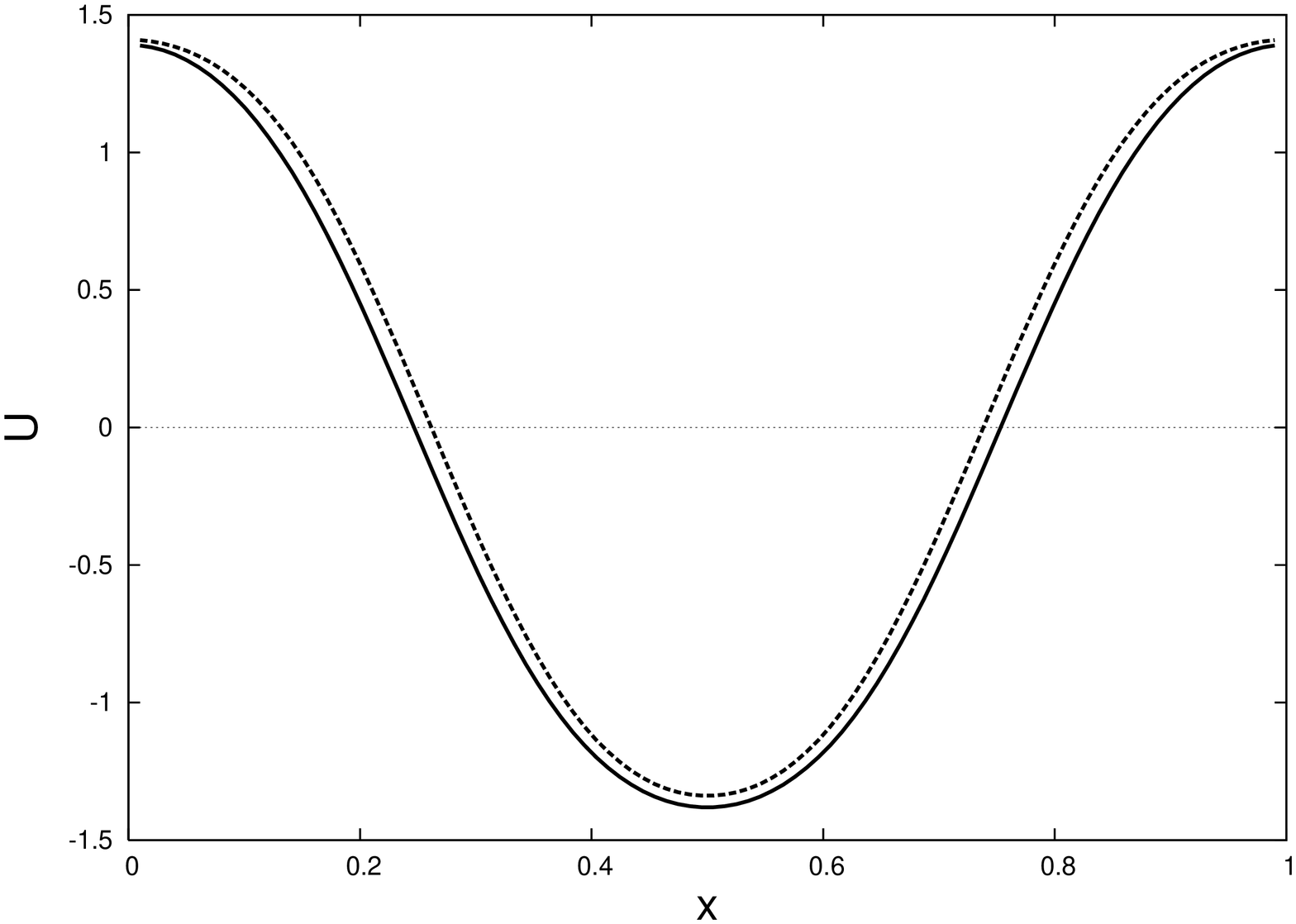}}
\resizebox{2.5in}{!}{\includegraphics[angle=-90]{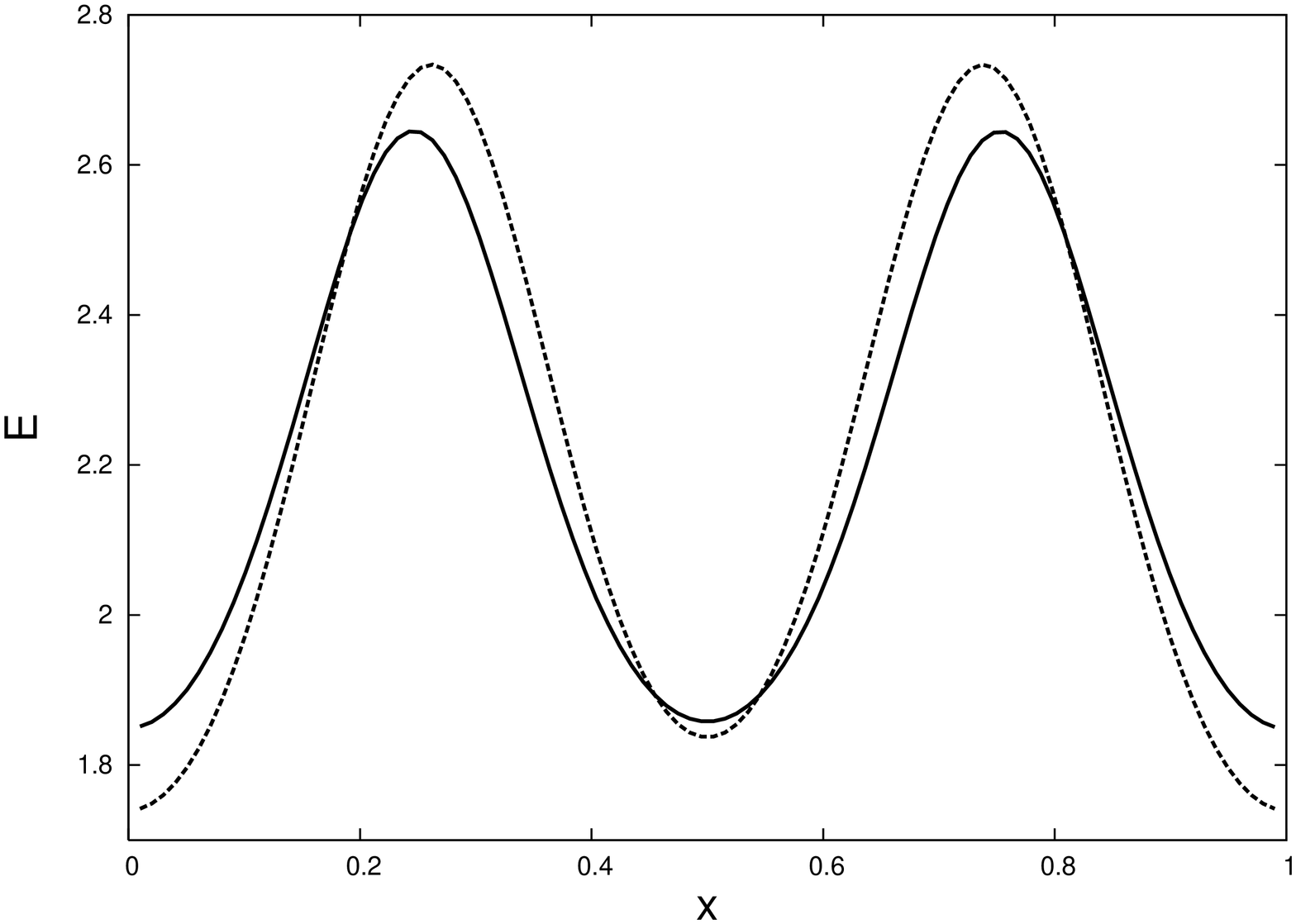}}
\resizebox{2.5in}{!}{\includegraphics[angle=-90]{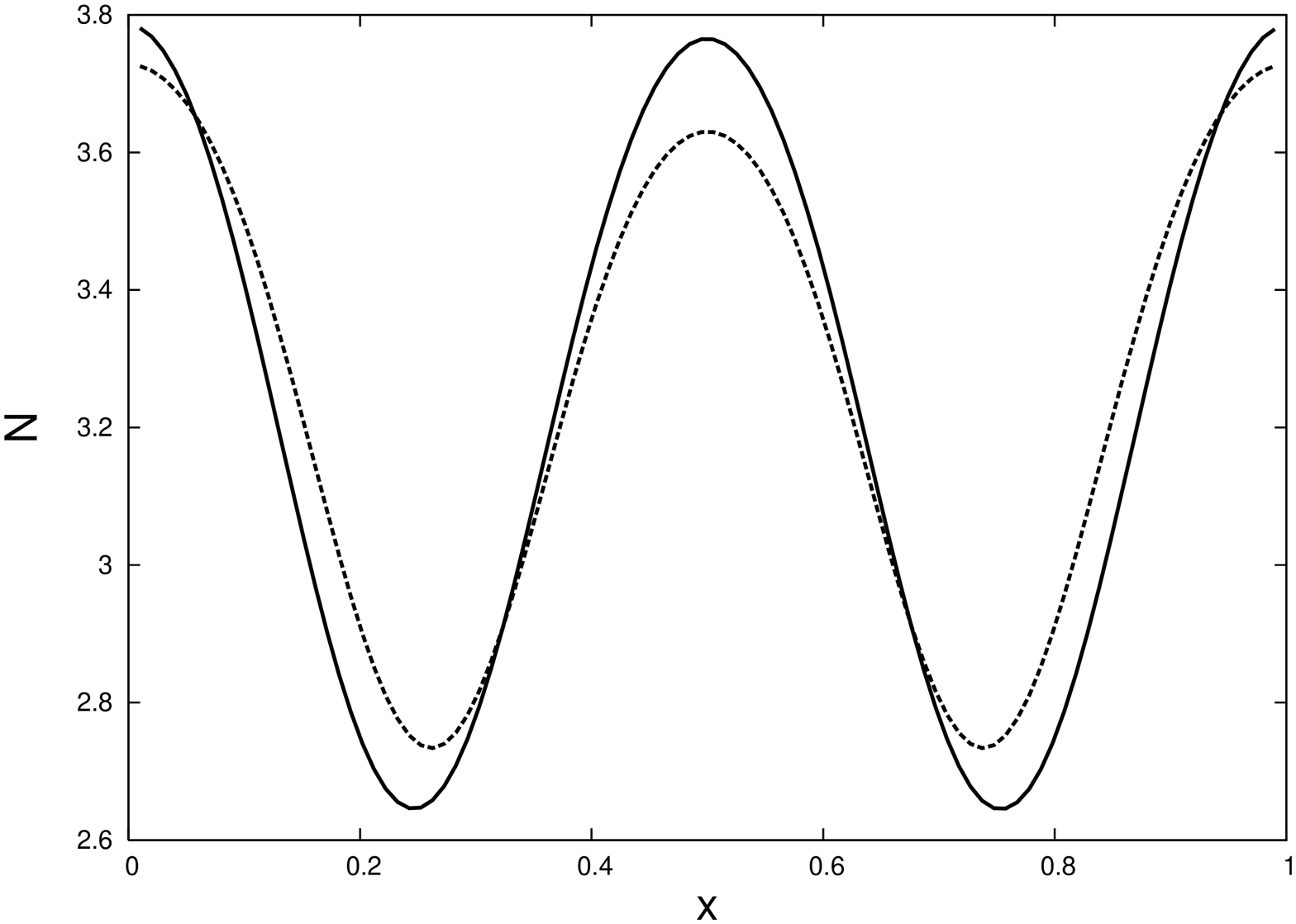}}
\end{center}
\caption{Stationary solutions computed from the integration of the
original model (solid) and from the reduced equation
(\ref{eq:redeqcollisional}) (dashed). The values of the parameters are
$\alpha_3=10^{-2}$, $D_2=10^{-4}$, $D_3=10^{-2}$, $D_0=D_1=D_4=0$,
$\mu_1=1$, $\mu_2=0$, $\Gamma=0.07$. The critical point is
$\Gamma_c=0.041853$.}
\label{fig:profilesCollision}
\end{figure}

The time-evolution predicted by this equation is compared to the
original model in Fig. \ref{fig:evolutionU0collision} for
$\delta=0.274$. The dynamics is very sensitive to $\delta$ and
Eq. (\ref{eq:redeqcollisional}) ceases to describe it accurately for
larger values of the expansion parameter. However, it is remarkable
that even for values of $\delta$ of order $1$, the stationary
solutions of the reduced equation give good approximations of the
exact ones (see Fig. \ref{fig:profilesCollision}, where
$\delta=0.82$).

\subsection{Collisional diffusion}

\begin{figure}[H]
\begin{center}
\resizebox{5in}{!}{\includegraphics[angle=-90]{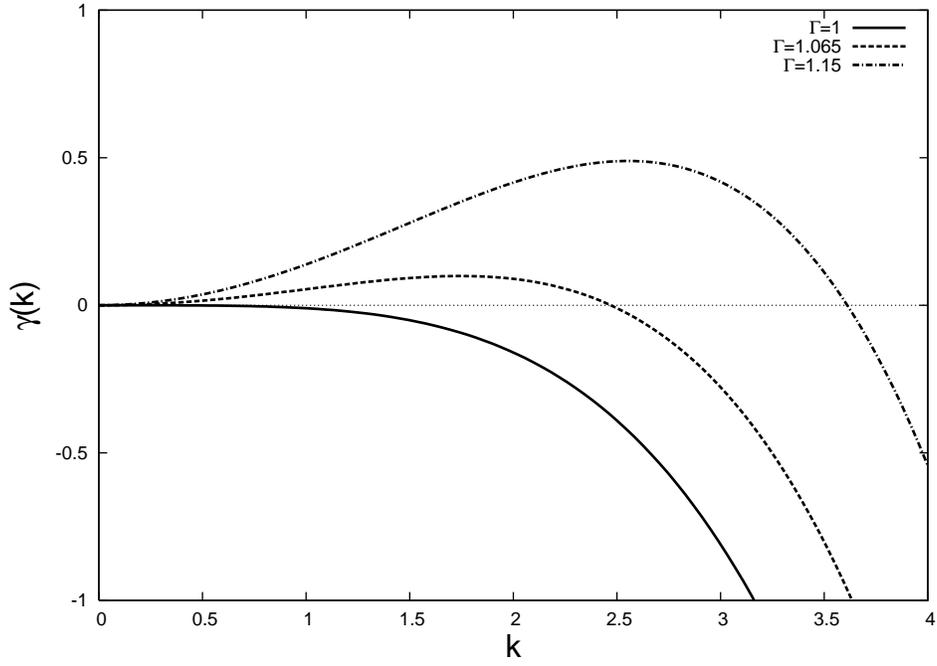}}
\end{center}
\caption{$\gamma(k)$ as a function of $k$ near the critical point. The
values of the parameters are $\alpha_3=10^{-2}$, $D_2=10^{-4}$,
$D_3=10^{-2}$, $D_0=D_1=D_4=0$, $\mu_1=0$, $\mu_2=1$. The critical
point is $\Gamma_c=1.1095$.}
\label{fig:grate_mu1zero}
\end{figure}

In this section we take $\mu_1=0$ and $\mu_2\neq 0$. In this case
\begin{equation}
\gamma(k)=(-\mu_2+\alpha_3E_f^2-D_2E_f^2k^2)k^2
\end{equation}
and the critical point is given by
\begin{equation}
E_c=\sqrt{\frac{\mu_2}{\alpha_3-k_c^2D_2}}\ .
\end{equation}
As we already know, if $\alpha_3/D_2<\pi^2$ there are no unstable
modes. Unlike the case of collisional drag, if $\alpha_3/D_2>\pi^2$
the most unstable (i.e. critical) mode is always $k=\pi$ (see
Fig. \ref{fig:grate_mu1zero}). In addition, for typical values of
$\alpha_3$ and $D_2$, $\alpha_3\gg\pi^2D_2$, and we can use the
approximation
\begin{equation}\label{eq:appcritpoint}
E_c=\sqrt{\frac{\mu_2}{\alpha_3}}\ .
\end{equation}

We are now ready to derive a reduced equation for the dynamics of
weakly unstable states in this case. The expansion of
(\ref{eq:1dmodelSigma}) in powers of $\delta$ yields:
\begin{equation}\label{eq:redeqstep1}
\delta^5\partial_{\bar\tau}U_1=
\delta^3\partial_{\bar\eta}^2\big[(\mu_2-\alpha_3E_c^2)U_1-
2\delta^2\alpha_3E_cE_2U_1-\delta^2D_2E_c^2\partial_{\bar\eta}^2U_1\big].
\end{equation}
Noting that the first term on the right-hand side vanishes if we use
the approximation (\ref{eq:appcritpoint}), we are left with
\begin{equation}
\partial_{\bar\tau}U_1= -\partial_{\bar\eta}^2\big[
2\alpha_3E_cE_2U_1+D_2E_c^2\partial_{\bar\eta}^2U_1\big].
\end{equation}

\begin{figure}[H]
\begin{center}
\resizebox{5in}{!}{\includegraphics[angle=-90]{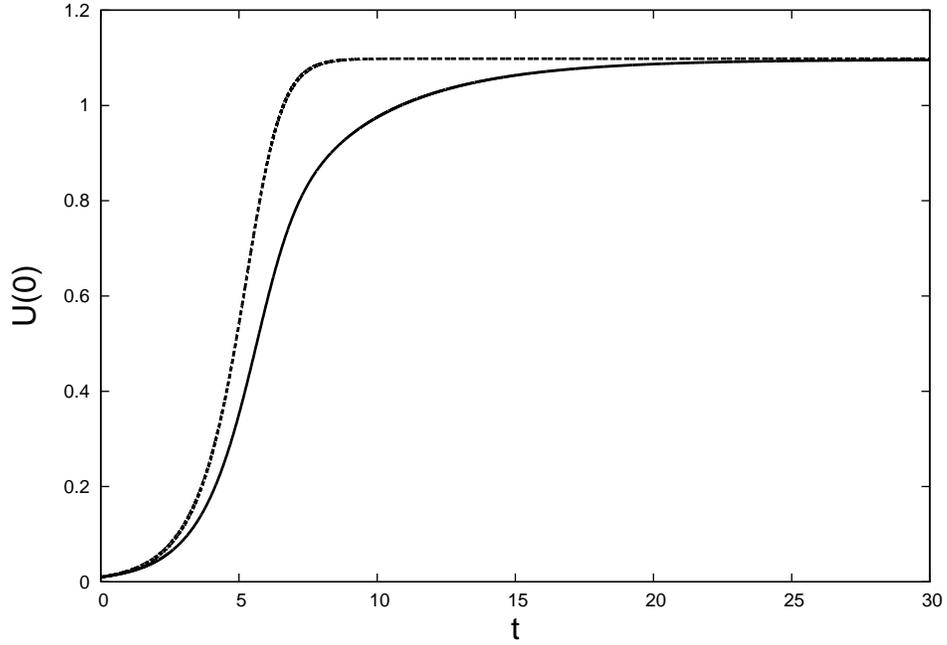}}
\end{center}
\caption{Time-evolution of $U(0)$ computed from
Eq. (\ref{eq:approxEqdiff}) (dashed) and from the integration of the
original model (solid). The values of the parameters are
$\alpha_3=10^{-2}$, $D_2=10^{-4}$, $D_3=10^{-2}$, $D_0=D_1=D_4=0$,
$\mu_2=1$, $\mu_1=0$, $\Gamma=1.2$. The critical point is
$\Gamma_c=1.1095$.}
\label{fig:evolutionU0Diff}
\end{figure}

\begin{figure}[H]
\begin{center}
\resizebox{5in}{!}{\includegraphics[angle=-90]{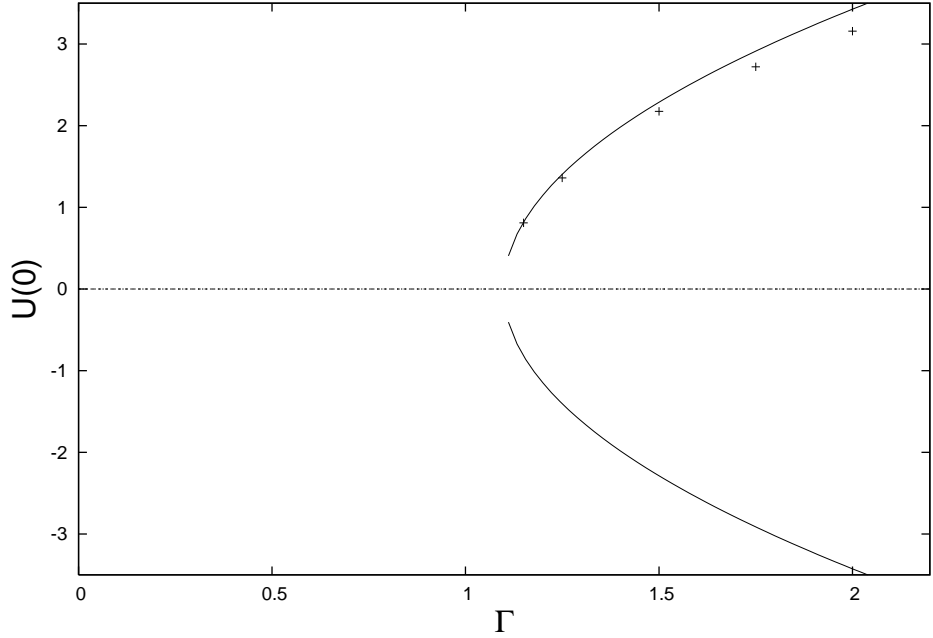}}
\end{center}
\caption{Structure of the bifurcation at the critical point. The
values of the parameters are $\alpha_3=10^{-2}$, $D_2=10^{-4}$,
$D_3=10^{-2}$, $D_0=D_1=D_4=0$, $\mu_2=1$, $\mu_1=0$. The critical
point is $\Gamma_c=1.1095$. The solid curve corresponds to the
analytical solutions of Eq. \ref{eq:stationaryGL}. The points have
been computed from the numerical integration of the original model. As
we can see, the reduced equation gives very accurate results even at
values of $\delta$ of order one.}
\label{fig:bifurcationDiff}
\end{figure}

\begin{figure}[H]
\begin{center}
\resizebox{2.5in}{!}{\includegraphics[angle=-90]{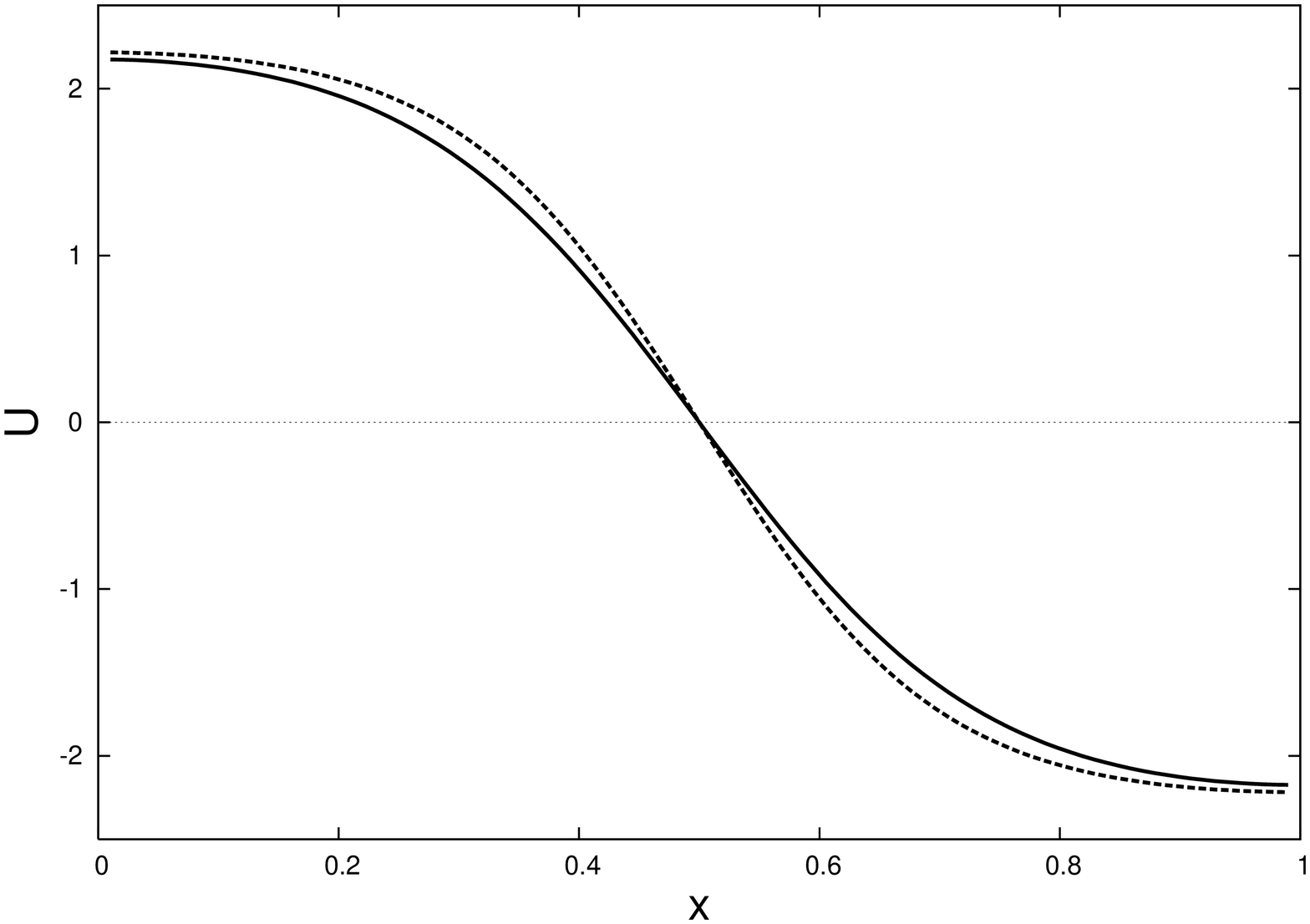}}
\resizebox{2.5in}{!}{\includegraphics[angle=-90]{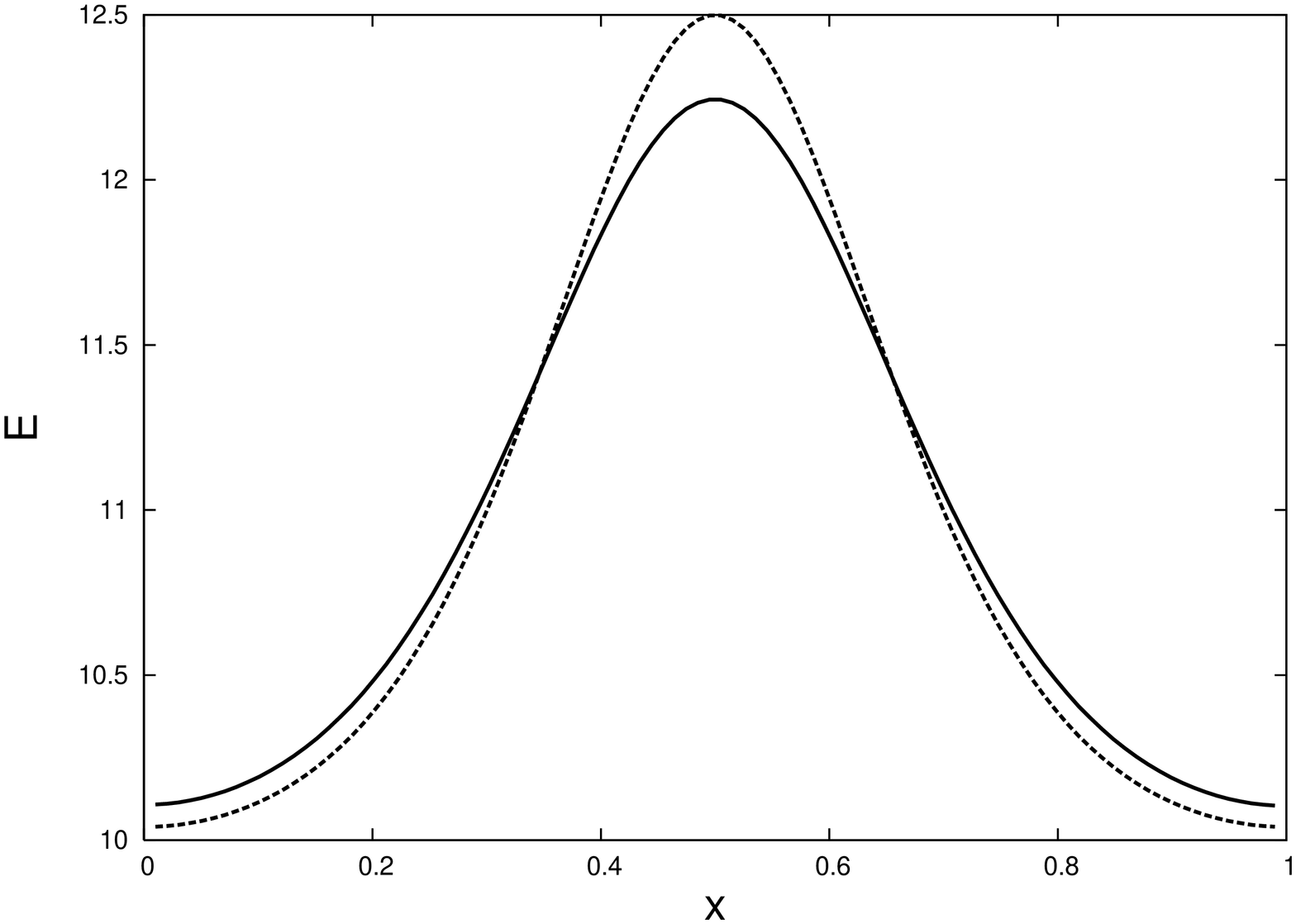}}
\resizebox{2.5in}{!}{\includegraphics[angle=-90]{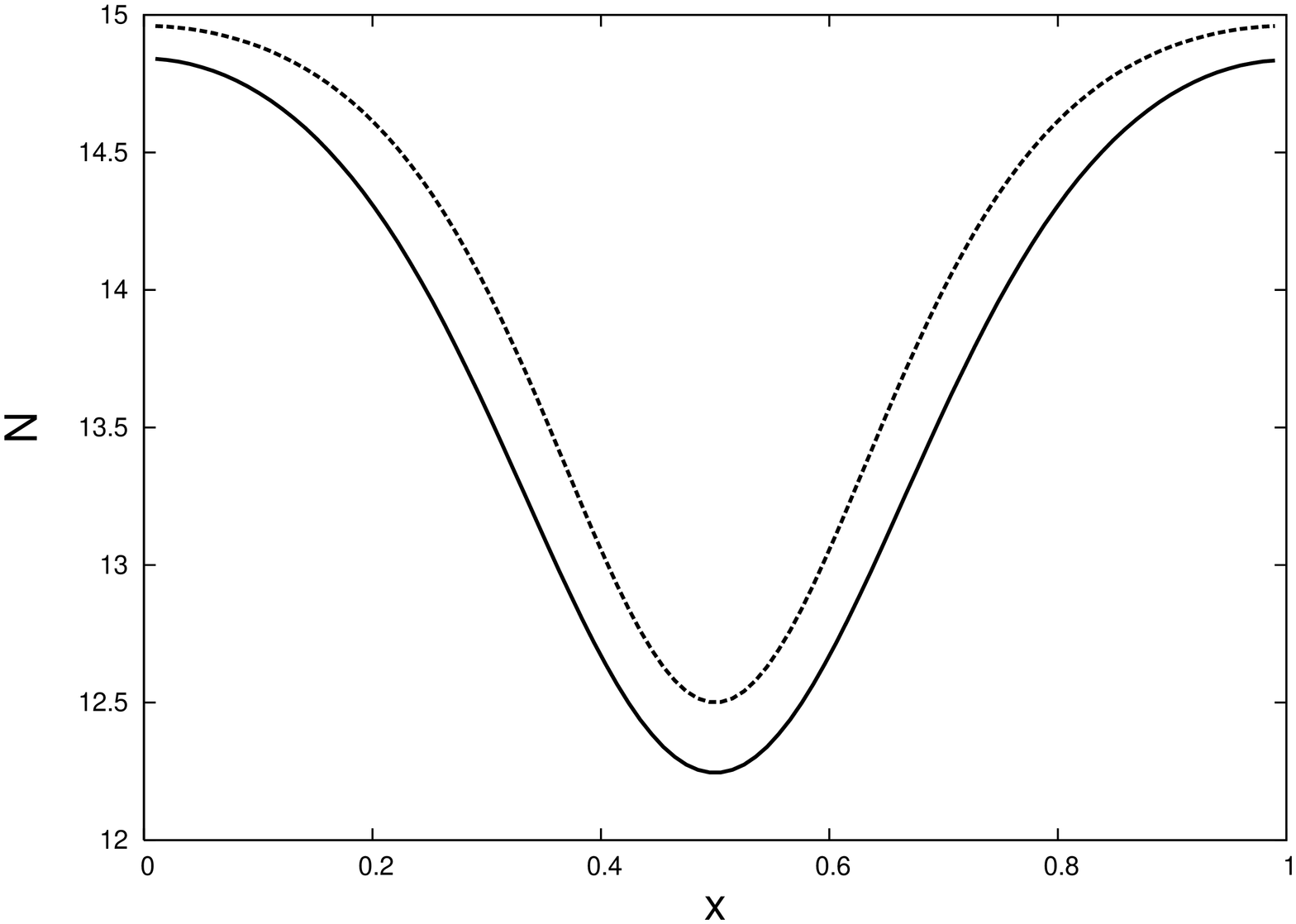}}
\end{center}
\caption{Stationary solutions computed from the integration of the
original model (solid) and from the GL equation
(\ref{eq:stationaryGL}) (dashed). The values of the parameters are
$\alpha_3=10^{-2}$, $D_2=10^{-4}$, $D_3=10^{-2}$, $D_0=D_1=D_4=0$,
$\mu_2=1$, $\mu_1=0$, $\Gamma=1.5$. The critical point is
$\Gamma_c=1.1095$.}
\label{fig:profilesDiff}
\end{figure}

Finally, we use (\ref{eq:slavingE}) and perform one more change of
variables:
\begin{equation}
\sigma=\frac{1}{\sqrt{E_c}}U_1,\quad \tau= \frac{1}{D_2}
\left(\frac{2\alpha_3E_c(D_4+D_3E_c)}{D_4+2D_3E_c}\right)^2\bar\tau,\quad
\eta=\sqrt{\frac{2\alpha_3(D_4+D_3E_c)}{D_2(D_4+2D_3E_c)}}\ \bar\eta
\end{equation}
obtaining the definitive form of the equation describing the weakly
non-linear dynamics of $U$:
\begin{equation}\label{eq:approxEqdiff}
\partial_{\tau}\sigma= -\partial_{\eta}^2\big[
\sigma-\sigma^3+\partial_{\eta}^2\sigma\big].
\end{equation}
In Fig. \ref{fig:evolutionU0Diff} we show a comparison between the
dynamics of Eq. (\ref{eq:approxEqdiff}) and that of the original
model, whereas Figs. \ref{fig:bifurcationDiff} and
\ref{fig:profilesDiff} are comparisons of the stationary solutions.

A remark is in order at this point. A consequence of using the
approximation (\ref{eq:appcritpoint}) in Eq. (\ref{eq:redeqstep1}) is
that the linear growth rate is modified. An easy calculation shows
that Eq. (\ref{eq:approxEqdiff}) gives the actual linear growth rate
(at order $\delta^2$) only if we make the replacement
\begin{equation}
\delta^2\mapsto \delta^2+\frac{D_2\pi^2}{\alpha_3}
\end{equation}
which is natural, since the accuracy of our approximation depends on
the quotient ${D_2\pi^2}/{\alpha_3}$. Thus, very close to the critical
point one has to take into account the above correction of $\delta$
when comparing the results of (\ref{eq:approxEqdiff}) with those
coming from the integration of the exact equations of the model.

Using that at the boundary, $\partial_x\sigma=\partial_x^3\sigma=0$,
and taking an initial condition such that $\int^1_0 U(x)\dd x=0$, the
stationary solutions of Eq. (\ref{eq:approxEqdiff}) are exactly the
solutions of:
\begin{equation}\label{eq:stationaryGL}
\sigma-\sigma^3+\partial_{\eta}^2\sigma=0,
\end{equation}
which is the time-independent Ginzburg-Landau equation for
second-order phase transitions. It is the same equation as in the case
of non-momentum-conserving Reynolds stress with a collisional drag
(see Ref. \cite{delCasCarLyn02}). We follow the lines of
Ref. \cite{delCasCarLyn02} to obtain analytical expressions for the
stationary solutions. The key observation is that there exists a
`conserved quantity'. Namely,
\begin{equation}
\frac{1}{2}(\partial_\eta\sigma)^2+\frac{\sigma^2}{2}\left(1-
\frac{\sigma^2}{2}\right)=\C,\ \C\in\mathbb R.
\end{equation}
This allows to reduce the solutions to quadratures:
\begin{equation}
\int\frac{\dd\sigma}{\sqrt{(\sigma^2-b_+)(\sigma^2-b_-)}}=
\int\frac{\dd\eta}{\sqrt{2}}
\end{equation}
with $b_{\pm}=1\pm\sqrt{1-4\C}$. Defining $m=b_-/b_+$ and
performing the change of variable $\sigma=\sqrt{b_-}\sin\phi$ we can
recast the solutions in the following form:
\begin{equation}
\sqrt{\frac{b_+}{2}}\eta=
\int_0^\phi\frac{\dd\theta}{\sqrt{1-m\sin^2\theta}}-A,
\end{equation}
where $A$ is an integration constant.

Hence, the solution for $\sigma(\eta)$ may be expressed as
\begin{equation}
\sigma(\eta)=
\sqrt{b_-}\mbox{sn}\left(\sqrt{\frac{b_+}{2}}\eta+A\big|m\right)
\end{equation}
where $\mbox{sn}(y|m)$ stands for the Jacobi elliptic function, which
is periodic in $y$. Its period is $P=4K(m)$, with
\begin{equation}
K(m)=4\int_0^{\pi/4}\frac{\dd\theta}{\sqrt{1-m\sin^2\theta}}
\end{equation}
the complete elliptic integral of the first kind. In order to write
explicitly the solution in terms of the original variable $U$ and
coordinate $x$, recall that
\begin{equation}
U=\delta \sqrt{E_c}\sigma,\quad
\eta=\delta\sqrt{\frac{2\alpha_3(D_4+D_3E_c)}{D_2(D_4+2D_3E_c)}} x \ .
\end{equation}
Then,
\begin{equation}
U(x)=\delta\sqrt{E_c b_-}\  \mbox{sn}\left(
\sqrt{\frac{b_+\alpha_3(D_4+D_3E_c)}{D_2(D_4+2D_3E_c)}}\ \delta
x+A\big|m\right).
\end{equation}
Thus, the wavelength of the solutions is
\begin{equation}
\lambda=\frac{4K(m)}{\delta}
\sqrt{\frac{D_2(D_4+2D_3E_c)}{b_+\alpha_3(D_4+D_3E_c)}}\ .
\end{equation}

The boundary conditions determine the integration constants $\C$
and $A$. In particular, they imply the quantization condition
\begin{equation}
n\frac{\lambda}{2}=1,\ 0\leq n\in\mathbb Z.
\end{equation}

\section{Conclusions and further work}

We have introduced a one-dimensional version of the phase transition
model considered in Ref. \cite{CarGarPedHid06} including a Reynolds
stress term with manifest momentum conservation. The model consists of
three envelope equations for the fluctuation level $E$, the poloidal
shear flow $U$ and the density gradient $N$. It possesses a critical
point corresponding to a second order transition whose natural control
parameter is $\Gamma$, the particle flux. Below the critical value,
$\Gamma_c$, the model has a non-trivial fixed point with
$U=0$. Through a linear stability analysis we have shown that if
$\Gamma>\Gamma_c$ the stationary solutions have non-zero shear flow
and reduced turbulent fluctuations. We have also studied the dynamics
of the model near (and above) the critical point. Defining a suitable
expansion around the critical point we have derived slaving conditions
for $E$ and $N$, so that they are determined from the value of
$U$. Then, we have found reduced equations for the weakly non-linear
dynamics of $U$. In the case of diffusive shear flow damping the
reduced equation is related to the Ginzburg-Landau equation, which
allows us to work out analytical expressions for the stationay
solutions of weakly unstable states.

The analysis performed in this work shows some interesting
differences with respect to previous one-dimensional versions of the
model in which momentum conservation is not implemented (see
Ref. \cite{delCasCarLyn02}), the most relevant of them concerning the
nature of the fixed points and the instabilities.

In Ref. \cite{delCasCarLyn02} the most unstable mode is always $k=0$,
which is related to the fact that in the model studied therein there
exist two (non-trivial) fixed points, one of them with $U=0$ and the
other one with $U\neq 0$. The introduction of a momentum-conserving
Reynolds stress makes the fixed point with non-zero $U$ disappear, so
that all stationary solutions with non-zero shear flow have
non-trivial spatial structure. This is connected to the results
derived in previous sections regarding the linear stability analysis
of the critical point. In the model with momentum conservation $k=0$
is always stable and the discussion of the structure of the most
unstable (critical) mode is more complicated. In general, the critical
mode depends on the values of the parameters. The presence of non-zero
critical mode is in agreement with the experimental findings reported
in Ref. \cite{PedHidCal05}.

A next step in the development of the model would be to incorporate
the diamagnetic term in order to study the L to H transition. In the
L-mode regime the model must incorporate the transport mesoscale,
which can be achieved by formulating transport equations in terms of
fractional derivative operators \cite{CarLynZas}. Such modification of
the model will lead to the dynamics of a reaction-diffusion system
\cite{delCasCarLyn03}. These issues will be addressed in future
publications.

\noindent{\bf Acknowledgements:} The authors acknowledge useful
discussions with L. Garc\'{\i}a, D. del Castillo Negrete,
F. Castej\'on and J. M. Reynolds. Part of this work has been sponsored
by the Association EURATOM-CIEMAT. I. C. gratefully acknowledges the
hospitality of ORNL during the final stages of this work.

\newpage

\newpage

\newpage

\newpage

\newpage

\newpage

\newpage

\newpage

\end{document}